\documentclass[prd,twocolumn,showpacs]{revtex4}
\usepackage{hyperref}
\usepackage{epsfig}
\usepackage{url}
\usepackage{amsmath}
\usepackage{amssymb}
\usepackage[pdftex]{color}

\begin{document}

\title{Halo abundances and shear in void models}
\author{David Alonso$^{1,2}$\thanks{E-mail: david.alonso@uam.es},
        \ Juan Garc\'ia-Bellido$^{1,2}$,
        \ Troels Haugb\o lle$^{3}$,
        \ Alexander Knebe$^{2}$}
\affiliation{ $^{1}$Instituto de F\'isica Te\'orica UAM-CSIC,
              Universidad Aut\'onoma de Madrid, 28049 Cantoblanco, Spain\\
              $^{2}$Departamento de F\'isica Te\'orica, Facultad de Ciencias, 
              Universidad Aut\'onoma de Madrid, 28049 Cantoblanco, Spain\\
              $^{3}$Centre for Star and Planet Formation, Natural History Museum of Denmark,
              University of Copenhagen,\\ \O ster Voldgade 5-7, DK-1350 Copenhagen, Denmark\\}

\begin{abstract}
We study the non-linear gravitational collapse of dark matter into halos through numerical N-body simulations of Lemaitre-Tolman-Bondi void models. We extend the halo mass function formalism to these models in a consistent way. This extension not only compares well with the simulated data at all times and radii, but it also
gives interesting clues about the impact of the background shear on the growth
of perturbations. Our results give hints about the possibility of constraining
the background shear via cluster number counts, which could then give rise
to strong constraints on general inhomogeneous models, of any scale.
\end{abstract}
\pacs{98.80.-k\hspace{\stretch{1}}IFT-UAM/CSIC-12-31}

\date{June 8, 2012}

\maketitle

\section{Introduction}\label{sec:intro}

Lema\^itre-Tolman-Bondi (LTB) void models have been proposed as a viable
alternative to dark energy. In these models one considers the possibility that
we might live inside a large underdense region (a void) and that the apparent
accelerated expansion of the Universe is only due to a misinterpretation of
the observations in terms of a homogeneous background
\citep{1998..ApJ..503..483,2000..AA..353..63,2001..MNRAS..326..287}
in which the expansion rate is the same everywhere. The coincidence problem of
$\Lambda$CDM (why now?) is substituted in these models by a violation of the
Copernican Principle (why here?), since, in order to accommodate the
observational constraints coming from the isotropy of the CMB and the matter
distribution, the position of the observer is restricted to be very close
($\sim1\%$) to the center of a highly spherical void. However, their ability
to explain away many evidences for dark energy without any dark component
has made LTB models a very attractive possibility. 

It has been shown
\citep{2006..PRD..73..083519,2007..JCAP..0702..019,2008..JCAP..0804..003} that
a gigaparsec-sized void can reproduce reasonably well the distance-redshift
relation deduced from current type Ia supernovae data. However, when these are
combined with other cosmological probes, LTB models run into trouble. In particular,
these models tend to predict a very high kSZ effect, due to the background
contribution \citep{2008..JCAP..0809..016,2011..CQG..28..164005,2011..PRL..107..041301}.
Also measurements of the local expansion rate combined with the full CMB power
spectrum seem to be incompatible, the former being too low in LTB models
\citep{2010..JCAP..1011..030,2011..PRD..83..103515}, and more recently it has 
been shown that the latest BAO and SNe-Ia data show some tension too
\citep{arXiv:1201.2790}. 

Nevertheless, research along these lines has been
very fruitful: these models have made us reconsider a non-standard approach to
cosmology, using the machinery developed around them we have been able to
consider observational effects due to the presence of large voids in a $\Lambda$CDM
cosmology \citep{2010..ApJ..718..1445,2011..PRD..84..083005,arXiv:1203.2180}, and
we now understand much better the evolution of perturbations in an inhomogeneous
background \citep{2008..PRD..78..043504,2009..JCAP..0906..025,arXiv:1202.1582}.

In a previous paper \citep{2010..PRD..82..123530} we presented the first
N-body simulations of LTB models, focusing on ensuring that the background
evolution was correctly reproduced. In the present work we have performed the
first study of halo statistics of one of the higher resolution simulations, and
give predictions regarding the non-linear accretion of dark matter halos in void
models. We will focus solely on the mass function and present a simple
modification to the Press-Schechter theory allowing to accommodate and explain
the effects introduced by using an LTB void model.

The main interest of this approach is not just the characterization of halo abundances
in LTB models, but a means to potentially distinguish whether the background space-time
is FRW or not. Here dark matter halos are acting as probes of the growth of density
perturbations and their mass function is extremely sensitive to the presence of a
finite background shear in large inhomogeneous voids of the LTB type. 

Moreover, we believe that smaller voids created due to the usual
non-linear gravitational collapse associated with the cosmic web must also induce
similar effects in the halo mass function, although at much smaller scales. If the
approach used in the present work were applicable to these smaller voids, the reported
contribution from the background shear could be included in the study of environmental
effects on halo formation \citep{2004..ApJ..605..1,2009..MNRAS..394..2109}.

\section{Theory}\label{sec:theory}

\subsection{The LTB metric}\label{ssec:GBH}

The Lema\^itre-Tolman-Bondi metric describes spaces with maximally symmetric
(spherical) 2-dimensional surfaces, and is given by
\begin{equation}
 ds^2=-dt^2+\frac{A'^2(t,r)}{1-k(r)}dr^2+A^2(t,r)d\Omega^2,
\end{equation}
for a matter source with negligible pressure and no anisotropic stress
($T^{\mu}_{\,\,\,\nu}=-\rho_M(t,r)\delta^{\mu}_0\delta^0_{\nu}$). The function
$A(t,r)$ acts as an $r$-dependent scale factor. It is easy to see that in this
framework the rates of expansion in the longitudinal ($r$) and transverse 
($\theta,\,\phi$) directions are, in general, different ($H_T\equiv\dot{A}/A$,
$H_L=\dot{A}'/A'$). With this setup, the Einstein equations can be written
as an effective Friedmann equation for a fixed $r$:
\begin{equation}
 H_T^2(t,r)=H_0^2(r)\left[\Omega(r)\frac{A_0^3(r)}{A^3(t,r)}+
(1-\Omega(r))\frac{A_0^2(r)}{A^2(t,r)}\right],
\end{equation}
where $A_0(r)\equiv A(t_0,r)$ can be gauged to $A_0(r)\equiv r$,
$H_0(r)\equiv H_T(t_0,r)$ and $\Omega(r)$ is the ratio between the average
matter density inside a sphere of radius $r$ and the critical density at
that radius, and acts as an effective $r$-dependent matter parameter.

The density profile of our simulated void follows the constrained-GBH model
\citep{2008..JCAP..0804..003}. In it the free function $\Omega(r)$ is
parametrized by the central underdensity $\Omega_{\rm in}$, the void radius
$r_0$ and the width of the transition void-background $\Delta r/r_0$. The 
other free function $H_0(r)$ is fixed by requiring a homogeneous Big Bang.
This means that the void is a pure growing mode that disappears at very high
redshift~\citep{2008..PRD..78..043504}. We fix the underlying
Friedmann-Robertson-Walker cosmology outside the void to be Einstein-de Sitter.
For a more thorough discussion of the LTB metric and the constrained-GBH model,
we refer the reader to \citep{2007..JCAP..0702..019,2008..JCAP..0804..003}.

\subsection{Linear perturbations and shear}\label{ssec:pert}

\begin{figure}
\centering
\includegraphics[width=0.45\textwidth]{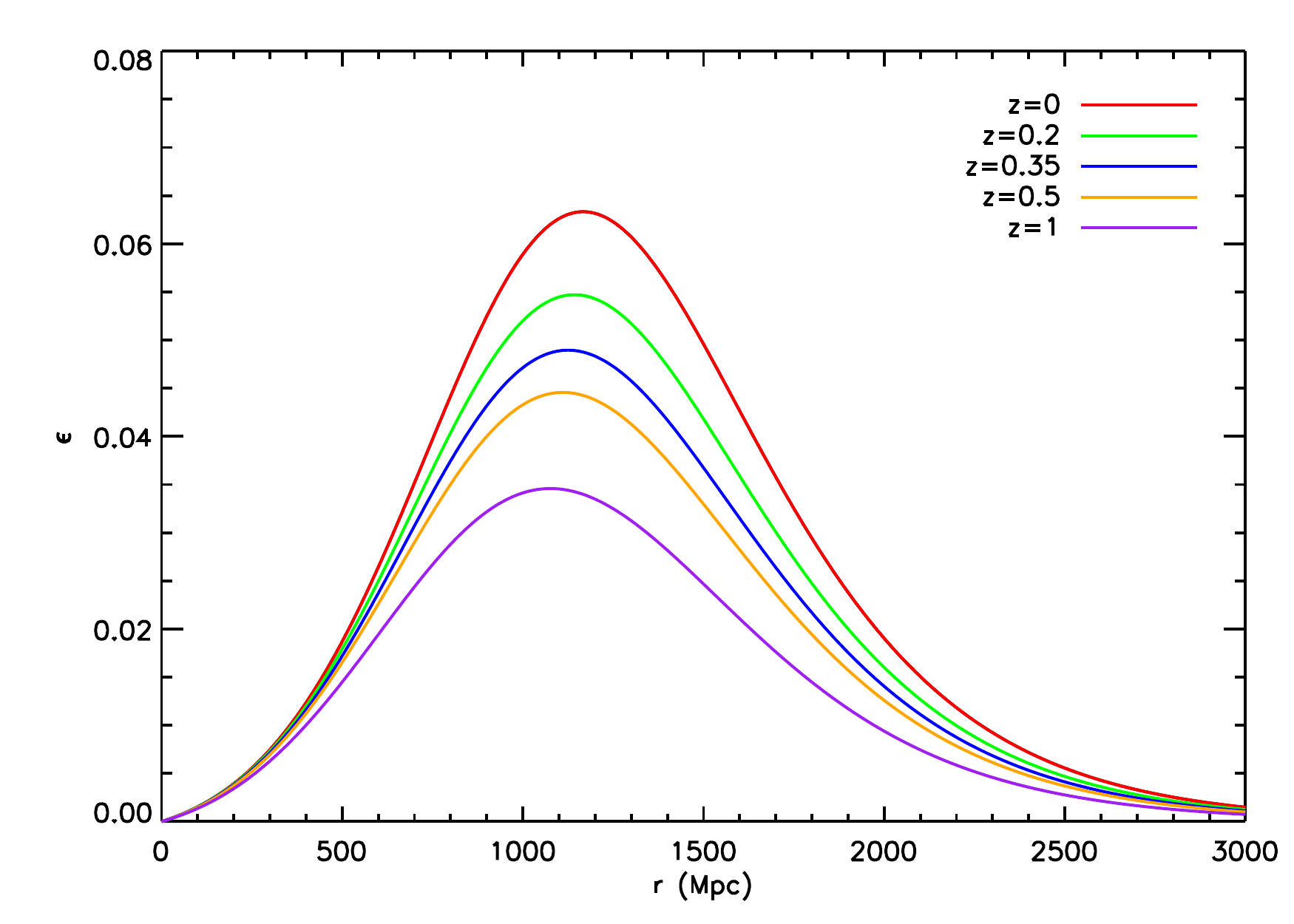}
\caption{Normalized shear parameter $\epsilon$ for the simulated LTB model.}
\label{fig:LTB_SHR}
\end{figure}

Although the equations describing perturbation theory in LTB models exist 
\citep{2009..JCAP..0906..025}, we do not yet have a good understanding of
their implications. The main problem arises
from the background not being homogeneous, and therefore the
perturbations cannot be split into irreducible representations of $SO(3)$ with
decoupled equations. In particular, the scalar modes couple to the vector and tensor
modes through the background shear tensor. The background shear can be
quantified as a normalized shear-to-expansion ratio 
\begin{equation}
\epsilon\equiv\sqrt{\frac{2}{3}\frac{\Sigma^2}{\Theta^2}}
=\frac{H_T-H_L}{2\,H_T+H_L}\,, 
\end{equation}
where $\Sigma^2=\Sigma_{ij}\Sigma^{ij}$ is the square of the background shear,
and $\Theta$ the expansion parameter of
a congruence of comoving geodesics (see \citep{2009..JCAP..0909..028} for
further details). For voids of practical interests this shear parameter
is usually small (in particular for the simulated model $\epsilon\lesssim0.06$,
see fig. \ref{fig:LTB_SHR}). Hence one would think that background shear effects
can be neglected, in which case the equations for the density
perturbations reduce, at a fixed radius $r$, to those of an FRW universe with
the corresponding effective cosmological parameters at that $r$. We write
this solution as
\begin{equation}\label{eq:pert_sol_0}
 \delta_0(t,r)\propto\,D(\Omega(r),A(t,r)/A_0(r)),
\end{equation}
where
\begin{align}\nonumber
 D(\Omega,a)&\equiv\frac{5}{2}\Omega\,h^2\,H(\Omega,a)\int_0^a
             \frac{da'}{[a'\,H(\Omega,a')]^3} \\
            &=a\cdot{}_2F_1\left[1,2;\frac{7}{2};\frac{\Omega-1}{\Omega}\,a\right]
 \end{align}
is the growth factor in an Open CDM universe with matter parameter $\Omega$, 
with $_2F_1(a,b;c;z)$ the Gauss hypergeometric function. The full density perturbation
equation has the form~\citep{Padmanabhan1996,2009..JCAP..0906..025}
\begin{equation}
\ddot\delta + 2H_T\dot\delta + (4\dot H_T + 6H_T^2)\delta = {\cal O}\left(\Sigma^2,\,\delta^2\right)\,.
\end{equation}
In the small shear limit, $\Sigma \to 0$, we propose the following Ansatz as an
approximate solution
\begin{equation}\label{eq:pert_sol_1}
 \delta_\alpha(t,r)= \delta_0(t,r) \,\Big(1+\alpha\,\epsilon(t,\,r)\Big),
\end{equation}
where the parameter $\alpha$ could, in principle, depend on $(t,\,r)$ and
possibly also on the cosmological parameters. 

\subsection{The mass function in LTB models}\label{ssec:mf}
Most of the information about the non-linear accretion of dark matter halos is
encoded in the mass function $n(M)\,dM$: the comoving number density of halos
with mass $M\in(M,M+dM)$. The first theoretical description of
the mass function was developed by Press and Schechter \citep{1974..ApJ..187..424}
(PS hereon) and
later re-derived and extended by Bond et al. in the so-called excursion set
formalism \citep{1991..ApJ..379..440}. Within this framework the abundance of
halos can be predicted as the abundance of points in space in which the linear
density contrast $\delta$ smoothed over a scale corresponding to the mass $M$
has crossed the spherical collapse threshold $\delta_c=1.686$. Although the PS
prediction describes qualitatively well the mass function, it fails to
reproduce its details (overpredicting the density of low mass objects and
underpredicting massive ones). Nevertheless it is a remarkable achievement
that one can estimate the abundance of non-linear structures using only linear
perturbation theory and the assumption that $\delta$ is gaussian-distributed.
The PS formula has been perfected using ellipsoidal collapse and
empirical parametrizations \citep{1999..MNRAS..308..119,2008..ApJ..688..709},
so that $n(M)$ can be calculated to very good accuracy, often using one
of the main results from this formalism: the mass function should be a universal
(cosmology-independent) function of the variance of the linear density contrast
field $\sigma(M,z)$ \citep{2000..MNRAS..321..372}. Here we will use
\begin{equation}\label{eq:massfunc}
 n(M,z)=\frac{\rho_M}{M}g(\sigma)\left|\frac{d\,\ln\,\sigma}{dM}\right|,
\end{equation}
where $\sigma\equiv\sigma(M,z)$ and $g(\sigma)$ is given by \citep{2007..MNRAS..379..1067}:
\begin{align}\nonumber
 &g(\sigma)\equiv\frac{a\,b\,\nu^b+2\,c\,\nu^2(1+a\,\nu^b)}{(1+a\,\nu^b)^2}\exp(-c\,\nu^2),\\\nonumber
 &(a,b,c)=(1.529,0.704,0.412), \,\,\,\nu\equiv\delta_c/\sigma.
\end{align}
We have also tried other parametrizations of the mass function 
\citep{1999..MNRAS..308..119,2008..ApJ..688..709} and checked that our results
did not depend significantly on this choice.

We follow the same rationale in order to calculate the mass function of halos
at a given $r$ and $t$ in an LTB model: since the simulated void arises from a
purely growing mode (i.e.: the Big Bang time is homogeneous), and
perturbations grow in a self-similar fashion, it is reasonable to assume that,
in order to calculate the variance of $\delta_M$ at $(t,r)$, 
we should rescale the variance $\sigma_{\rm out}(M,z)$ of the density
perturbations outside the void, by a factor
\begin{equation}
 f(t,r)=\frac{\delta_\alpha(t,r)}{\delta_\alpha(t,r\rightarrow\infty)}\,,
\end{equation}
where the density contrast is computed theoretically according to Eq.~(\ref{eq:pert_sol_1}), 
and evaluated in Fig.~\ref{fig:LTB_MF},  with and without the shear correction.
Thus, our model for the mass function $n(M,z,r)$ at a given radius $r$ is
Eq.~(\ref{eq:massfunc}) with $\sigma(M,z)$ substituted by $\sigma_{\rm out}(M,z)
\,f(t,r)$.

Note however, that our main results will be quoted in terms of the cumulative
mass function within a sphere of radius $r$ centered at the origin of the LTB patch:
\begin{equation}
 n(>M,<r,z)\equiv\frac{3}{4\pi\,r^3}\int_0^rr'^2dr'\int_M^{\infty}dM' \,n(M',r',z),
\end{equation}
since this observable has better statistics.

\section{The simulation and the halo catalog}\label{sec:methods}
A more detailed description of the simulation we have used can be found in
\citep{2010..PRD..82..123530} (simulation $\mathcal{H}$). It has $960^3$
particles in a box of size $L=2400$ Mpc $h^{-1}$, which sets the mass
resolution to $m_p=4.2\times10^{12}\,M_{\odot}\,h^{-1}$. The simulated void
has a size of $r_0=1100$ Mpc, a transition width of $\Delta r/r_0=0.3$
and an underdensity of $\Omega_{\rm in}=0.25$ (see section \ref{ssec:GBH}).
The background cosmology is Einstein-de Sitter ($\Omega=1$) with $h_{\rm out}=
0.43$. The small-scale perturbations are set using a power spectrum with
$n_s=1$ and $\sigma_8=0.9$. It must be noted that an LTB void with these
parameters is in fact ruled out, since the supernovae and baryon acoustic 
oscillation data seems to be mutually in conflict given a particular LTB 
profile~\citep{arXiv:1201.2790}. We only use this simulation with a large 
background shear as a toy model to test our Ansatz about the mass function. 
The technique used to simulate LTB voids is based
on setting the initial conditions (IC) appropriately by modifying the IC
generator to take into account the large-scale perturbation induced by the
presence of the void. This must be done at a high enough redshift so that the
void can be regarded as a linear perturbation.  A modified version of the
{\tt 2LPT} code \citep{2006..MNRAS..373..369} was used for this stage.
Once the ICs are set, they are plugged into {\tt Gadget2}
\citep{2005..MNRAS..364..1105}, which we run in pure tree-mode.

The halo catalog has been extracted using the AMIGA halo finder {\tt AHF} 
\citep{2009..ApJS..182..608, 2004MNRAS.351..399G}. It maps the particle
content to an adaptively smoothed density field and locates the position
of possible halos as local overdensities. Once the gravitationally bound
particles around these have been extracted, the extent and mass of each
halo is computed as:
\begin{equation}
 M(r_{\rm vir})=\frac{4\pi}{3}r_{\rm vir}^3\rho_c\,\Delta,
\end{equation}
where we have used $\Delta=200$ as a collapse threshold \citep[e.g.][]{2011MNRAS.415.2293K}.

\section{Results}\label{sec:results}

\begin{figure}
\centering
\includegraphics[width=0.45\textwidth]{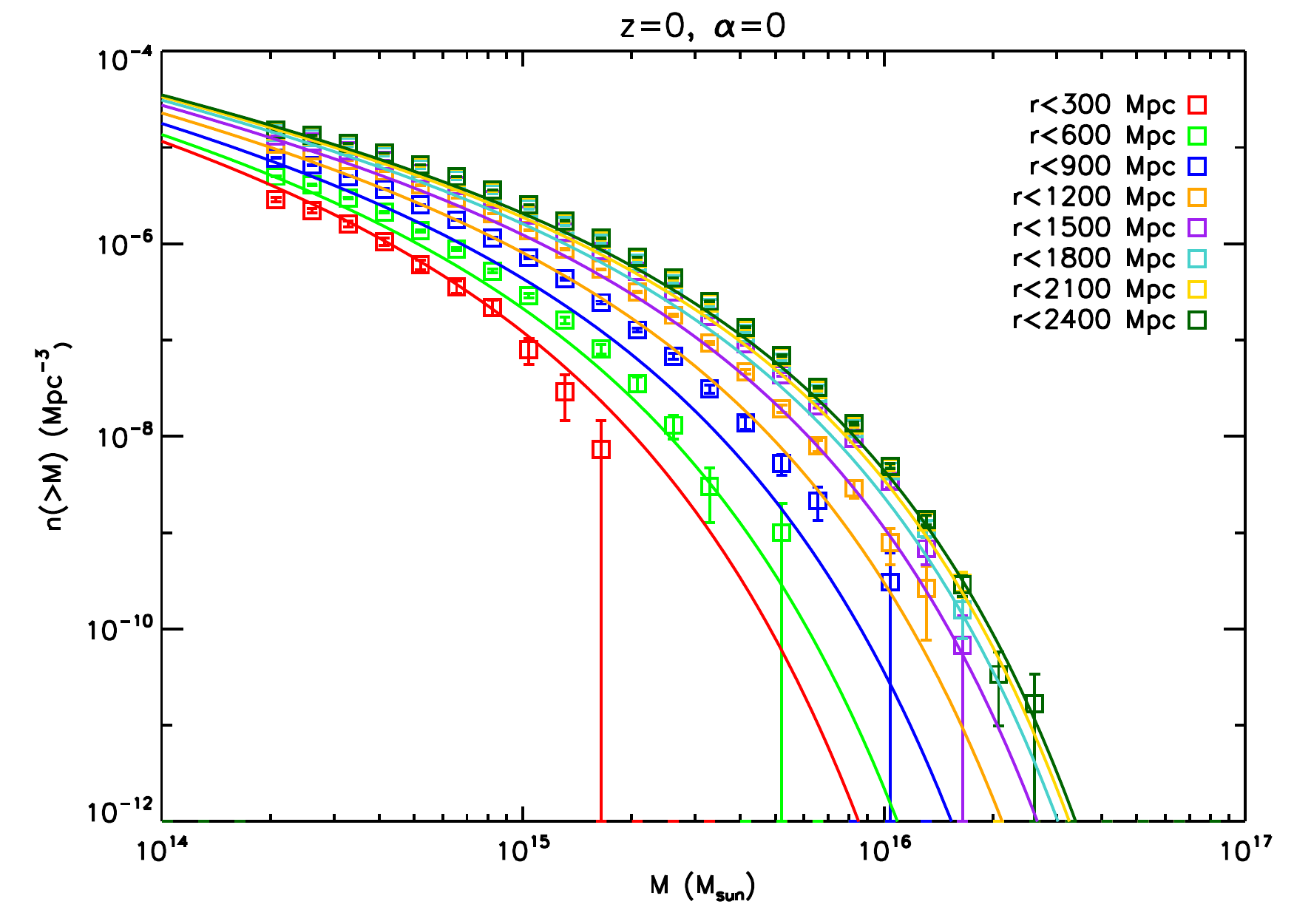}
\includegraphics[width=0.45\textwidth]{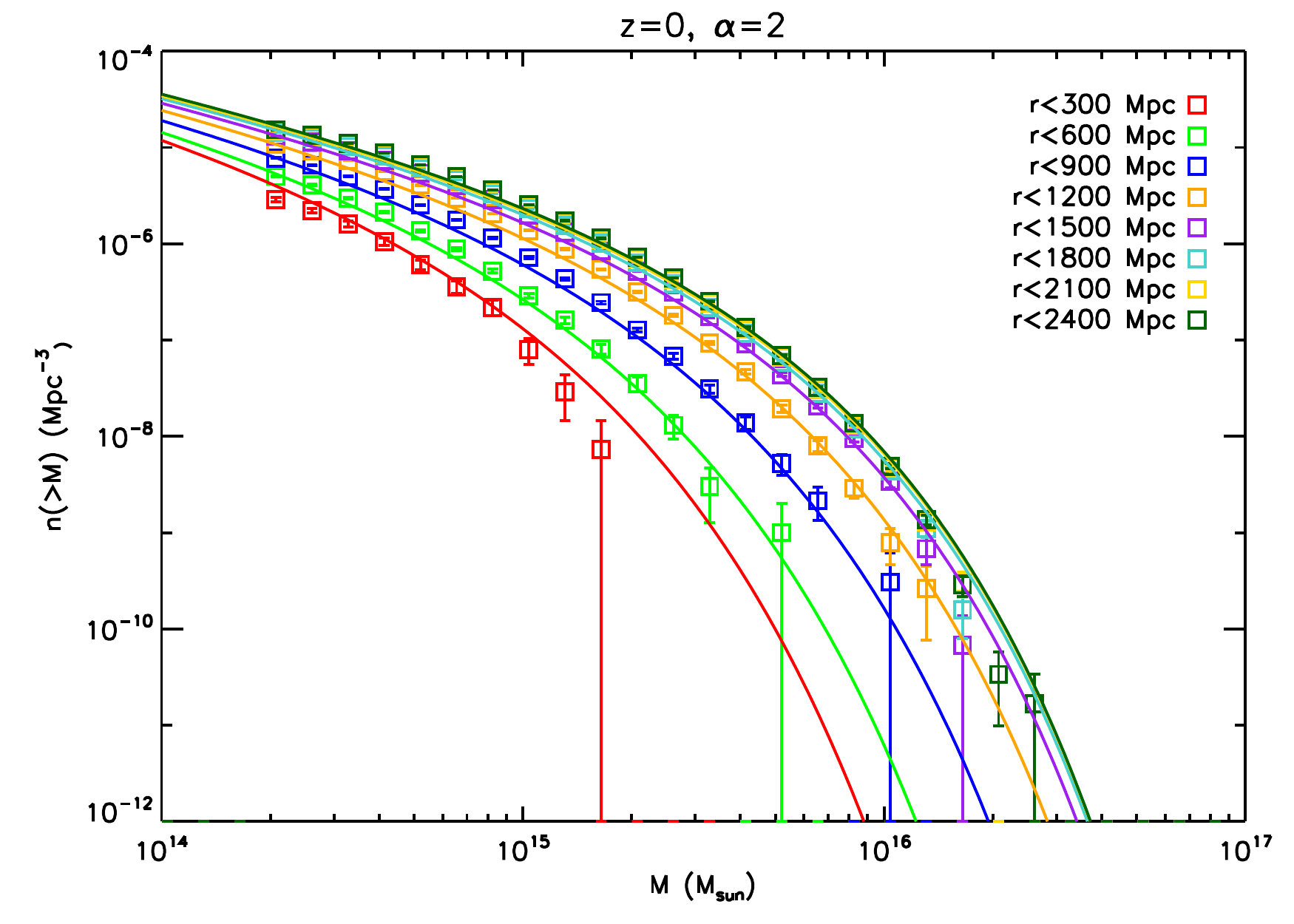}
\caption{Cumulative mass function for halos inside spheres of different radii
compared with the theoretical prediction without the background shear correction
($\alpha=0$, top panel) and with a first order shear correction, where
$\alpha=2\,\forall\,r,\,t$ (bottom panel).}
\label{fig:LTB_MF}
\end{figure}

\begin{table}
\begin{center}
\begin{tabular}{c|c|c}
$r$ (Mpc)& $\xi_N^2 (\alpha=0)$ & $\xi_N^2 (\alpha=2)$ \\
\hline
300  &  0.093 & 0.112 \\
600  &  0.682 & 0.104 \\
900  &  8.321 & 0.173 \\
1200 &  2.394 & 0.047 \\
1500 &  1.474 & 0.072 \\
1800 &  0.478 & 0.050 \\
2100 &  0.177 & 0.032 \\
2400 &  0.095 & 0.061 \\
\hline \hline
\end{tabular} 
\end{center}
\caption{Goodness of fit of our theoretical approach with and without a
first order shear correction for different radii at $z=0$. Similar results hold at all other
redshifts (see table \ref{tab:gof2}).} \label{tab:gof1}
\end{table}

\begin{table}
\begin{center}
\begin{tabular}{c|c|c}
$z$& $\xi_N^2 (\alpha=0)$ & $\xi_N^2 (\alpha=2)$ \\
\hline
0    & 1.620 & 0.072 \\
0.2  & 1.163 & 0.065 \\
0.35 & 1.411 & 0.042 \\
0.5  & 0.667 & 0.047 \\
1    & 3.494 & 0.824 \\
\hline \hline
\end{tabular} 
\end{center}
\caption{Goodness of fit of our theoretical approach with and without a
1-st order shear correction for different redshifts (summing over all radii).} \label{tab:gof2}
\end{table}

\begin{figure}
\centering
\includegraphics[width=0.45\textwidth]{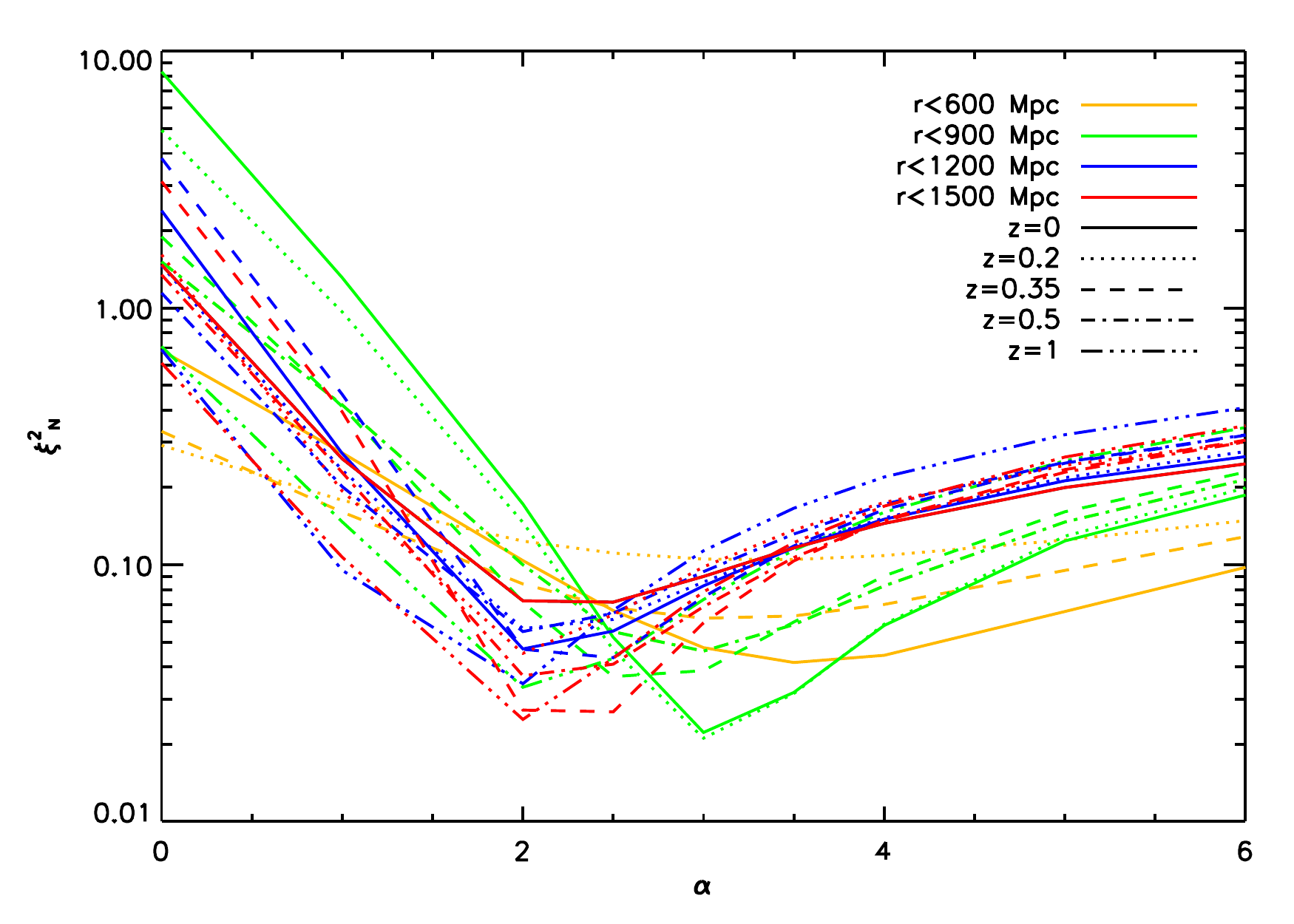}
\includegraphics[width=0.45\textwidth]{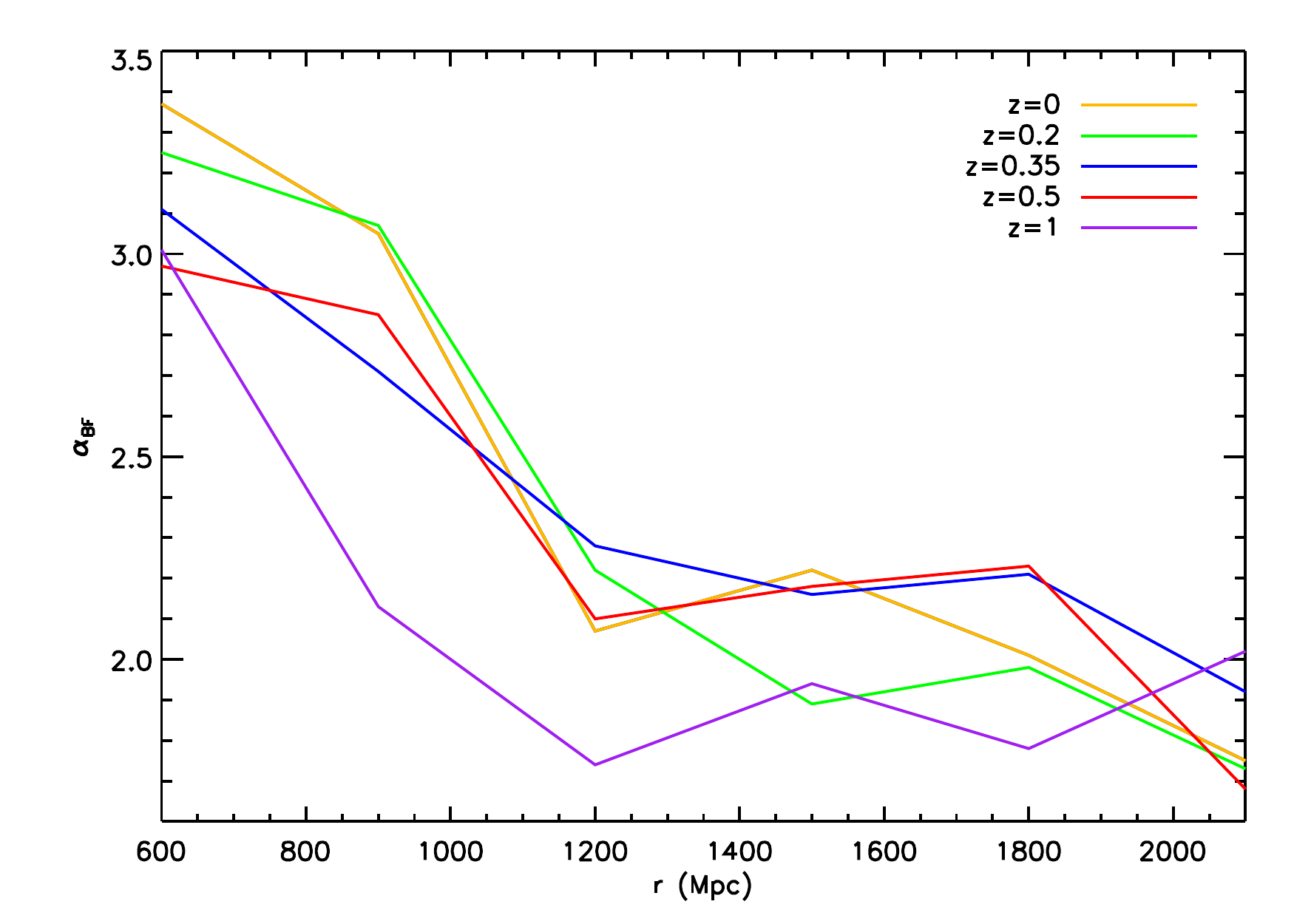}
\caption{Top panel: dependence of $\xi_N^2$ (eq. \ref{eq:xi2}) with the value
of the shear correction term $\alpha$ at different $r$ and $t$. Bottom panel: best
fit value of $\alpha$ for different radii and redshifts, showing a mild evolution with
$r$.}
\label{fig:xi2}
\end{figure}

Once the halos have been identified, we can compute $n(>M,<r,z)$ merely by
counting the number of halos with mass above $M$ inside a sphere of radius $r$
and dividing by the comoving volume of this sphere. The results are shown in
figure \ref{fig:LTB_MF}. It is easy to see (top panel) that, while
approximating the growth of perturbations by its 0-th order in $\epsilon$
(eq. \ref{eq:pert_sol_0}) yields a reasonably good fit at small and large
radii (where the background shear vanishes), it fails to reproduce the halo
abundances at intermediate radii. In the bottom panel we can see, however,
that adding a non-zero 1-st order correction (eq. \ref{eq:pert_sol_1}) solves
this problem. Furthermore, we have found that this correction seems to be
almost independent of $r$ and $t$, with $\alpha\sim2$.

We have quantified the goodness of fit of our approach (with and without the
shear correction term) using the measure:
\begin{equation}\label{eq:xi2}
 \xi_N^2(r,z)\equiv\frac{1}{N-1}\sum_{i}^N
  \left(\frac{n(>M_i,<r,z)-n_{i}(r)}{n(>M_i,<r,z)}\right)^2.
\end{equation}
Here $n(>M,<r,z)$ is given in section \ref{ssec:mf}, $n_{i}(r)$ is the cumulative
mass function obtained from the simulation for a mass $M_i$ within a sphere of
radius $r$, and $N$ is the number of mass bins. Figure \ref{fig:xi2} (top panel)
shows the value of $\xi_N^2$ for different choices of $\alpha$ at different $r$
and $z$. A value of $\alpha\sim2$, found as the median of the best-fit values
for all the calculated curves, gives a good fit in all cases with only a very mild
dependence on $r$ and $t$ (shown on the bottom panel of Fig. \ref{fig:xi2}). The
improvement due to the shear-correction term can also be seen in table
\ref{tab:gof1}, in which we have calculated the goodness of fit with and without
the shear correction for different radii at $z=0$. Table \ref{tab:gof2} shows the
same result for different redshifts summing over all radii. This improvement is
especially evident at intermediate radii, where $\epsilon$ is larger and therefore
its effects more important.

It would be extremely interesting to investigate whether and how the value of
$\alpha$ depends on the cosmological parameters: if this parameter turned out
to be independent of the void model, one should be able to predict its value
from some approximation in perturbation theory. However, this is work in
progress and we defer the presentation of it to future work that will also
make use of better resolved simulations.

\section{Discussion \& conclusions}\label{sec:discussion}

We have extracted the halo content from an LTB N-body simulation and analyzed
the halo abundances at different masses and radii. The main conclusions from
this study are:
\begin{itemize}
 \item The theoretical description of the halo mass function in FRW
       cosmologies can be fully extended to LTB void models by adding just one 
       parameter ($\alpha$) that accounts for the effect of the background
       shear on the evolution of matter density perturbations.
 \item The value of this parameter ($\sim2$) seems to be constant in time and
       only mildly dependent on the position in the void. Whether
       this value depends weakly on the void model is still work in progress.
\end{itemize}

If the shear correction turns out to be practically independent of the void model parameters,
halo abundances could potentially be used to constrain the amount of background
shear, a crucial test for general inhomogeneous cosmological models. A toy model that has 
been considered in the past in connection with LTB scenarios, and which could benefit from 
our analysis of background shear, is the swiss-cheese model~\citep{2007..PRD..76..123004,
2008..JCAP..0806..021,2008..PRD..77..023003}. We leave for the future such investigation.
It is also worthwhile exploring whether our results apply to voids of astrophysical
scales, of tens of Mpc, since they could have an effect on the modelling of
the environmental dependence of dark matter halo properties, as well as the 
backreaction of non-linear gravi\-tational collapse on the background evolution.

\section*{Acknowledgments}

The authors would like to thank the journal reviewers for their useful comments.
DAM acknowledges support from a JAE-Predoc contract. We also acknowledge 
financial support from the Madrid Regional Government (CAM) under the program 
HEPHACOS S2009/ESP-1473-02, from MICINN under grant  AYA2009-13936-C06-06 
and Consolider-Ingenio 2010 PAU (CSD2007-00060), as well as from the European 
Union Marie Curie Initial Training Network "UNILHC" PITN-GA-2009-237920.
AK is supported by the {\it Spanish Ministerio de Ciencia e Innovaci\'on} (MICINN)
in Spain through the Ramon y Cajal program as well as the grants AYA 2009-13875-C03-02,
AYA2009-12792-C03-03, CSD2009-00064, and CAM S2009/ESP-1496. TH is supported by the Centre
for Star and Planet Formation which is financed by the Danish National Science Foundation.
Computer time for the simulations was provided by the Danish Center for Scientific Computing (DCSC).

\label{lastpage}

\end{document}